\documentclass[twocolumn,prb,superscriptaddress]{revtex4-2}
\newcommand{\pagenumbaa}{1}
\usepackage{lipsum}	
\usepackage[english]{babel}
\usepackage{graphicx}
\usepackage{framed}
\usepackage{amsmath}
\usepackage{amsthm}
\usepackage{amssymb}
\usepackage{amsfonts}
\usepackage{enumerate}
\usepackage[utf8]{inputenc}
\usepackage{datetime}
\usepackage{indentfirst}
\usepackage{float}
\usepackage{physics}
\usepackage{siunitx}
\usepackage{transparent}
\usepackage{adjustbox}
\usepackage{colortbl}
\usepackage[table,dvipsnames]{xcolor}
\usepackage[colorlinks,citecolor=blue,linkcolor=blue,urlcolor=blue]{hyperref}
\usepackage[capitalize]{cleveref}

\begin{document}

%%%%%%%%%%%%%%%%%%%%%%%%%%%%%%%%%%%%%%%%%%%%%%%%%%%%%%%%%%%%%%%%%%%%
% Place your title here
%%%%%%%%%%%%%%%%%%%%%%%%%%%%%%%%%%%%%%%%%%%%%%%%%%%%%%%%%%%%%%%%%%%%
\title{
Charge-4e supercurrent in a two-dimensional InAs-Al superconductor-semiconductor heterostructure}

%%%%%%%%%%%%%%%%%%%%%%%%%%%%%%%%%%%%%%%%%%%%%%%%%%%%%%%%%%%%%%%%%%%%
% Place names of authors here
%%%%%%%%%%%%%%%%%%%%%%%%%%%%%%%%%%%%%%%%%%%%%%%%%%%%%%%%%%%%%%%%%%%%
\author{Carlo Ciaccia}
\email[E-mail: ]{Carlo.Ciaccia@unibas.ch}
\affiliation
{Quantum- and Nanoelectronics Lab, University of Basel, CH-4056 Basel, Switzerland}

\author{Roy Haller}
\affiliation
{Quantum- and Nanoelectronics Lab, University of Basel, CH-4056 Basel, Switzerland}

\author{Asbj\o{}rn C. C. Drachmann}
\affiliation
{Center for Quantum Devices, Niels Bohr Institute, University of Copenhagen, 2100 Copenhagen, Denmark}
\affiliation
{NNF Quantum Computing Programme, Niels Bohr Institute, University of Copenhagen, 2100 Copenhagen, Denmark}

\author{Tyler Lindemann}
\affiliation
{Department of Physics and Astronomy, Purdue University, West Lafayette, Indiana 47907, USA}
\affiliation
{Birck Nanotechnology Center, Purdue University, West Lafayette, Indiana 47907, USA}

\author{Michael J. Manfra}
\affiliation
{Department of Physics and Astronomy, Purdue University, West Lafayette, Indiana 47907, USA}
\affiliation
{Birck Nanotechnology Center, Purdue University, West Lafayette, Indiana 47907, USA}
\affiliation
{School of Electrical and Computer Engineering, Purdue University, West Lafayette, Indiana 47907, USA}
\affiliation
{School of Materials Engineering, Purdue University, West Lafayette, Indiana 47907, USA}

\author{Constantin Schrade}
\affiliation
{Center for Quantum Devices, Niels Bohr Institute, University of Copenhagen, 2100 Copenhagen, Denmark}

\author{Christian Sch\"onenberger}
\affiliation
{Quantum- and Nanoelectronics Lab, University of Basel, CH-4056 Basel, Switzerland}
\affiliation
{Swiss Nanoscience Institute, University of Basel, Klingelbergstrasse 82, Basel, Switzerland}

\begin{abstract}
%{\center \bf \noindent Abstract\\}
%\vspace{0.3cm}
\section*{Abstract}
\vspace{-0.3cm}
Superconducting qubits with intrinsic noise protection offer a promising approach to improve the coherence of quantum information. Crucial to such protected qubits is the encoding of the logical quantum states into wavefunctions with disjoint support. Such encoding can be achieved by a Josephson element with an unusual charge-4e supercurrent emerging from the coherent transfer of pairs of Cooper-pairs. In this work, we demonstrate the controlled conversion of a conventional charge-2e dominated to a charge-4e dominated supercurrent in a superconducting quantum interference device (SQUID) consisting of gate-tunable planar Josephson junctions. We investigate the ac Josephson effect of the SQUID and measure a dominant photon emission at twice the fundamental Josephson frequency together with a doubling of the number of Shapiro steps, both consistent with the appearance of charge-4e supercurrent. Our results present a step towards novel protected superconducting qubits based on superconductor-semiconductor hybrid materials.
\end{abstract}

\maketitle

\setcounter{page}{\pagenumbaa}
\thispagestyle{plain}

\section{Introduction}
\label{sec:Introduction}
\vspace{-0.3cm}
\begin{figure*}[htp]
	\centering
	\includegraphics[width=\textwidth]{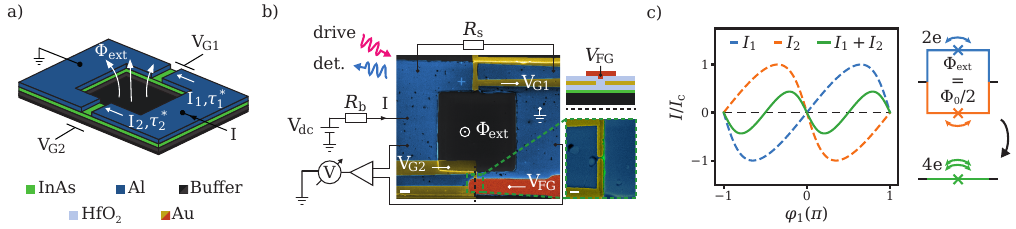}
	\caption{\textbf{Device description and measurement setup.} a) Circuit schematic of a dc superconducting quantum interference device (SQUID) formed by two gate tunable Josephson junctions with effective transmission probabilities $\tau^*_1$, $\tau^*_2$, threaded by the external flux $\Phi_{\mathrm{ext}}$. b) False color electron micrograph of the device and experimental setup. Each junction is fabricated by selectively removing the epi-Al layer (blue) over $\SI{150}{\nano\meter}$ long stripes. The charge carrier density in the exposed InAs two-dimensional electron gas (green) is tuned by a set of electrostatic gates ($V_{\rm G1}$, $V_{\rm G2}$ and $V_{\rm FG}$) shown in yellow and orange, which are galvanically isolated from the loop by $\SI{15}{\nano\meter}$ of HfO$ _2 $ (light blue). Dc and ac current bias are defined through the voltage drop over a bias resistor $ R_\text{b}= \SI{1}{\mega\ohm}$. The SQUID is shunted to ground with $ R_\text{s}= \SI{10}{\ohm}$. We send a microwave tone to the device, and also detect photon emission. The scale bar in the main figure is $\SI{1}{\micro\meter}$, and the scale bar in the zoom-in is $\SI{300}{\nano\meter}$.  c) Individual components $ I_1 $ (blue) and $ I_2 $ (orange) and total current (green) flowing through a symmetric SQUID as a function of the phase drop $ \varphi_1 $ at $\Phi_{\mathrm{ext}} = \Phi_0/2$. The current phase relation of both junctions is plotted using a single channel short diffusive model with an effective transparency $\tau^* = 0.86$. The current is normalized to units of the critical current $ I_{\rm c} $. The schematic of the SQUID helps visualizing the requirements for a $ \sin(2\varphi) $ junction: a dominant 4e supercurrent is obtained with a symmetric SQUID biased at $ \Phi_0/2 $.}
	\label{fig:Fig1}
\end{figure*}
The Josephson effect describes the dissipationless current flow between two weakly coupled superconductors. Today, numerous technologies are based on this fundamental quantum phenomenon, ranging from
superconducting qubit devices~\cite{Wallraff2004,Arute2019,Clarke2008,Castellanos2007,Krinner2022} to parametric amplifiers~\cite{frattini20173,frattini2018optimizing,miano2022frequency}.

Regardless of whether the weak link consists of an insulator or a normal conducting material, the supercurrent is a periodic function of the phase difference $ \varphi $ between the superconductors~\cite{Golubov2004}. In a Josephson tunnel junction, the supercurrent arises from the coherent tunnelling of individual Cooper-pairs through the insulating barrier, each carrying a charge 2e~\cite{JOSEPHSON1962251}. The current-phase relation (CPR) in this case is given by $ I(\varphi) = I_{\rm c}\sin(\varphi) $, with $ I_{\rm c} $ being the critical current. However, when the superconductors are separated by a conducting weak link, such as a semiconductor or a metal, coherent transport of multiple Cooper-pairs can also occur, resulting in a non-sinusoidal CPR~\cite{DellaRocca2007,Spanton2017,Nichele2020,endres2022currentphase,Stoutimore2018}. In general, the CPR of the junction can be expanded in a Fourier series as:
\begin{equation}
I(\varphi) = \sum_{m=1}^{\infty}c_m\sin(m\varphi).
\label{eqn:fourierCPR}
\end{equation}
The $ \sin(m\varphi) $ terms correspond to processes involving the simultaneous, coherent transport of $ m $ Cooper-pairs~\cite{Heikkil2002,Chauvin2006} carrying a charge $ m\times 2 $e. The amplitude of the higher harmonic terms $c_{m}$, $ m>1 $, reflects the probability of multi-Cooper-pair transport and decreases with higher harmonics, indicating that transport across the junction arises mainly from individual Cooper-pairs. Often, the CPR can be described by the junction transparency $ \tau $, defined as the transmission probability of electron in the weak link. The more transparent a junction is, the higher the ratio between successive Fourier coefficients $|c_{m+1}(\tau)/c_{m}(\tau)|$.

Several theoretical proposals~\cite{Ioffe2002,Dou2002,Brooks2013,groszkowski2018coherence,Paolo_2019,Schrade2022,Smith2020,Maiani2022,leroux2023cat} have investigated possible advantages of using a so-called $ \sin(2\varphi) $ Josephson junction (JJ) for the realization of a parity protected superconducting qubit. In this case, the parity of the Cooper pairs is protected by using a Josephson element with a dominant second harmonic term $ c_2 $ in Eq.\eqref{eqn:fourierCPR}, corresponding to the supercurrent being carried by pairs of Cooper pairs with charge 4e. The qubit states can be therefore encoded into the even and odd parity of the number of Cooper-pairs on a superconducting island.

Important steps towards realizing a parity protected qubit have been taken with superconducting quantum interference devices (SQUIDs) made of tunnel junctions arranged in a rhombus geometry~\cite{Gladchenko2009,Bell2014}. By designing the loop inductances and the junctions position, it possible to engineer a CPR with a large second harmonic component $|c_{2}/c_{1}|\sim 0.5$~\cite{Pop2008}, corresponding to an effecting transparency $ \tau^* \sim 1$~\cite{Heikkil2002}. When the magnetic flux through the SQUID is tuned to half a flux quantum $\Phi_0/2$, the first harmonic is suppressed due to destructive interference, leaving a dominant second harmonic term. This method relies on the fabrication of identical junctions, and departures from symmetry spoils parity protection.

A promising alternative approach is based on gate tunable hybrid superconducting-semiconducting materials with high transparency channels. In Ref.~\cite{Larsen2020} the authors realize a $ \sin(2\varphi) $ element with a SQUID made of proximitized InAs nanowires, where local gate control of each junction allows precise balancing of the first harmonics. They show that the qubit relaxation time increases by an order of magnitude when the qubit is tuned close to the protected regime. However, for practical use of the parity protected qubit, the Josephson energy of the second harmonics in the balanced configuration must be at the same time much larger than the residual Josephson energy coming from the first harmonics (for a long relaxation time) and much larger than the island charging energy (for small dephasing rate). The few conduction channels in the nanowires limit the maximum obtainable critical current and make the last requirement difficult to satisfy.

Hybrid two-dimensional materials have seen in recent years a great improvement in growth techniques that allow up-scaling and offer the advantage of wide gate tunability and top-down fabrication~\cite{Kjaergaard2016,Hendrickx2018}. In this work, we report the observation of a $4\text{e}$ supercurrent in a SQUID consisting of two planar Josephson junctions formed in an InAs two-dimensional electron gas proximitized by an epitaxial Al layer~\cite{Shabani2016,Lee2019}. Even if the operation of superconducting qubits has already been shown in this material platform~\cite{Casparis2018}, the realization of high quality resonators on III-V substrates remains a challenging task. Therefore, here we investigate the contribution of the 4e supercurrent by measuring the evolution in frequency of the ac Josephson radiation emitted by the SQUID as a function of a dc bias voltage. The high transparency of these JJs~\cite{Nichele2020} allows us to engineer an effective CPR in which the first harmonic is suppressed due to destructive interference, leaving a dominant second harmonic term. To achieve this, we balance the critical current of the junctions with local gate voltages and tune the magnetic flux through the SQUID loop to half a flux quantum $\Phi_0/2$. In the balanced configuration, radiation measurements reveal a pronounced suppression of emission at the fundamental Josephson frequency in favour of a strong ac signal at twice this frequency. We corroborate this finding by additionally detecting fractional half Shapiro steps, characteristic of a $ \sin(2\varphi) $ junction.

%-------------------------------------------------------------------------------------------------------------------
\section{Results and Discussion}
%-------------------------------------------------------------------------------------------------------------------
\subsection{Device and procedures}
\label{subsec:Setup}
\vspace{-0.3cm}
%-------------------------------------------------------------------------------------------------------------------
\begin{figure*}[htb]
	\centering
	\includegraphics[width=\textwidth]{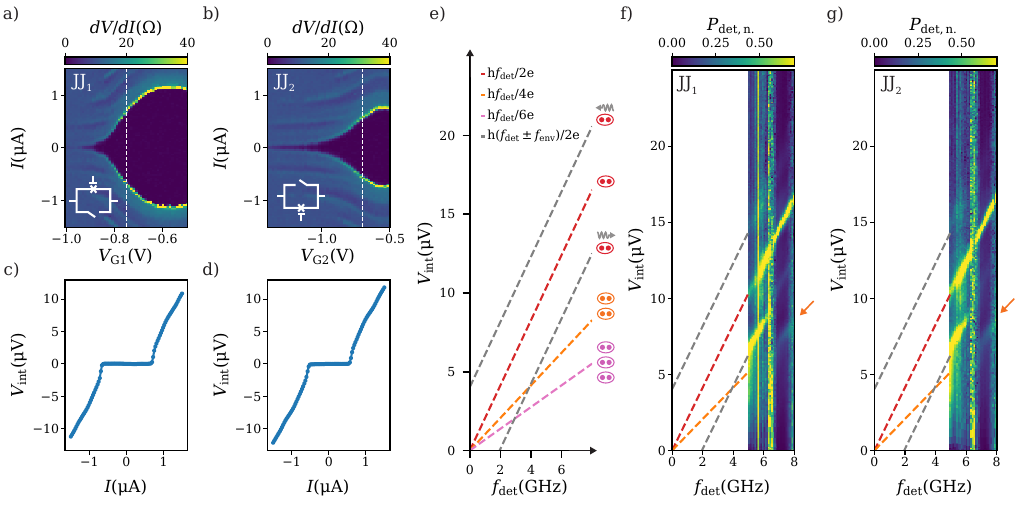}
	\caption{\textbf{Dc and ac Josephson effect of the individual junctions.} a) and b). Differential resistance $ dV/dI $ of the individual Josephson junctions JJ$ _1 $ and JJ$ _2 $ as a function of gate voltage $ V_{\rm G1} $,$V_{\rm G2}$ and current bias $ I $. c) and d). Integrated voltage $ V_{\rm int} $ as a function of current bias $ I $ at $V_{\rm G1}=-0.75$~V and $V_{\rm G2}=-0.7$~V obtained by integrating the corresponding $ dV/dI $ along the white dashed lines shown in a) and b). e) Illustration of the expected peak evolution in the emission spectrum of voltage biased JJ as a function of detection frequency $f _{\rm det} $. A junction with finite transparency emits photons at the fundamental Josephson frequency (red dashed line) and integer multiples of it (orange and pink dashed lines), here corresponding to the coherent transport of pairs of Cooper-pairs. The dashed grey lines indicate processes associated to the up- and down-conversion of environmental photons at frequency $f _{\rm env}$. f) and g) Normalized radiation power $P _{\rm det,norm} $ as a function of $f _{\rm det} $ and $ V_{\rm int}$ for the same configuration in c) and d). The orange arrow points to the $4$e emission peak.}
	\label{fig:Fig2}
\end{figure*}
A simplified schematic of the device is shown in Fig.~\ref{fig:Fig1}a). A superconducting loop, threaded by an external magnetic flux $ \Phi_{\mathrm{ext}} $, is interrupted on each arm by a section where the superconductor has been selectively removed. The Josephson junctions are formed in an InAs two-dimensional electron gas (green) which is proximitized by the close vicinity to an epitaxial Al layer (blue) grown on top. By locally removing the Al top layer with etching techniques that are detailed in the Methods section, we form InAs weak links. Local gate electrodes, $V_{\textrm{G1}}$ and $V_{\textrm{G2}}$, allow us to tune the electron density in the weak links and, consequently, adjust the critical currents of the JJs. The hereby formed Josephson junctions are symmetric by design, but the wet etching step produced two different widths: $\sim3\,\mu$m for JJ$ _1 $ and $\sim2.5\,\mu$m for JJ$ _2 $. Despite of fabrication-related asymmetries, we were still able to tune the SQUID into a symmetric configuration by leveraging the gate tunability of the semiconducting weak link. Junctions this wide typically contain many conduction channels with a bimodal distribution of transparency values distributed between zero and one~\cite{Kulik1975,Dorokhov1984,Nazarov1994,Beenakker1997}. Earlier experiments on the same material platform have shown that the CPR in these junctions can be described by a single channel short diffusive junction model~\cite{ciaccia2023gate,Nichele2020} with an effective transparency $ \tau^* \sim0.86 $.

Fig.~\ref{fig:Fig1}b) depicts a false-color electron micrograph of the device and the experimental setup. We apply a dc-current via the voltage drop over a bias resistor $R_{\rm b} = 1\,{\textrm{M}}\Omega$. We damp the SQUID with a shunt resistor $R_{\rm s} = 10~\Omega$ to enable a continuous transition from the superconducting to the normal conducting state. The $ 10~\si{\ohm} $-shunt increases the region of stable voltage drop across the junction, and at the same time it reduces both heating and hysteretic behaviours. The differential resistance is measured using standard lockin techniques. Furthermore, the microwave-setup allows probing the ac Josephson effect in two ways. On one hand, the Josephson radiation emitted from the SQUID under finite dc bias can be detected with a spectrum analyser. Second, the reverse experiment can be performed, namely, irradiating the device with a microwave tone and measuring its dc response.

Fig.~\ref{fig:Fig1}c) shows the interference between the supercurrent $ I_1 $ flowing in JJ$ _1 $ (blue dashed curve) and the supercurrent $ I_2 $ in JJ$ _2 $ (orange dashed curve) at $ \Phi_{\mathrm{ext}}=\Phi_{0}/2 $. The total supercurrent flowing through the SQUID (green solid curve) is:
\begin{equation}
I=I_1(\varphi_1,\tau^*_1)+ I_2(\varphi_2,\tau^*_2).
\label{eqn:I_SQUID}
\end{equation}
The phase drops over the two JJs are related by the fluxoid relation $ \varphi_1 - \varphi_2 = 2\pi\Phi_{\mathrm{ext}}/\Phi_{0} $. Here, we have assumed that the phase difference between the two JJs is solely given by the externally applied flux, neglecting loop and mutual inductances, which is justified in our device~\cite{ciaccia2023gate}. When the loop is flux biased at $ \Phi_{\mathrm{ext}}=\Phi_{0}/2 $ and the JJs are the same ($ \tau^*_1 = \tau^*_2 $), Cooper-pairs are transferred with the same amplitude but opposite phase through the SQUID arms, resulting in a destructive interference of the 2e contribution with periodicity $2\pi$ and in a constructive interference of the $4\text{e}$ supercurrent with periodicity $\pi$. In this way, it is possible to engineer an effective $ \sin(2\varphi) $ junction.
%-------------------------------------------------------------------------------------------------------------------
\subsection{Ac and dc Josephson effect from single junction}
\label{subsec:Radiation from a Single junction}
\vspace{-0.3cm}
%------------------------------------------------------------------------------------------------------------------
\begin{figure*}[htp]
	\centering
	\includegraphics[width=\textwidth]{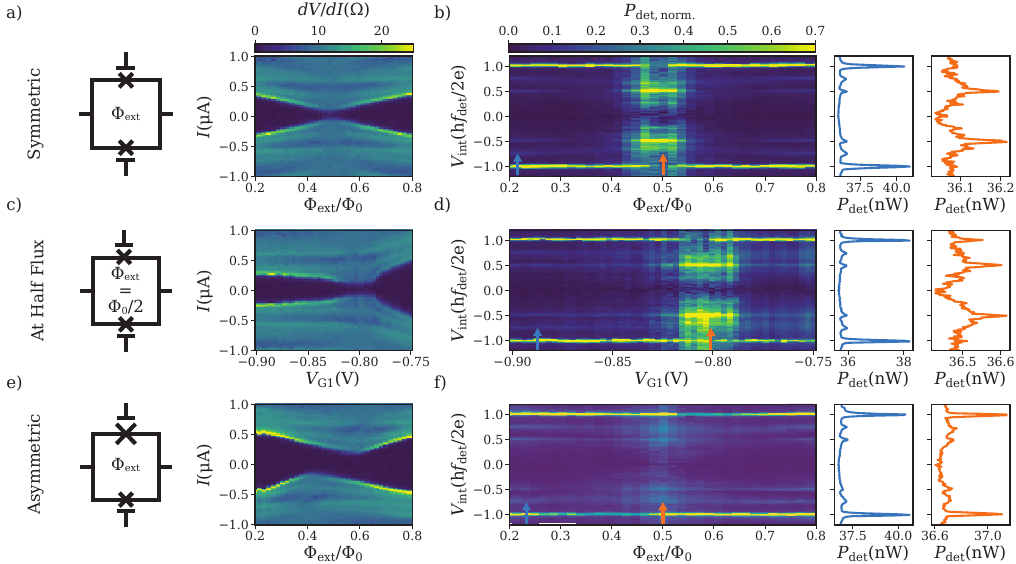}
	\caption{\textbf{Tuning a superconducting quantum interference device into a symmetric configuration.} a) Differential resistance $ dV/dI $ of the SQUID as a function of external flux $ \Phi_{\mathrm{ext}} $ and current bias $ I $ for symmetric junctions. Here, the gate voltage on one junction is $ V_{\rm G1} = -0.865 $~V and the gate voltage on the other junction is $ V_{\rm G2} = -0.9 $~V. In this balanced configuration, there is no diode effect. b) Normalized radiation power $P _{\rm det,norm.}$ at a detection frequency $ f_{\text{det}} = 7.1~$GHz plotted vs external flux $\Phi_{\mathrm{ext}}$ and normalized voltage drop over the SQUID $V _{\rm int}$. The map is measured as the same time as in a). At half flux quantum, the $2$e radiation signal is suppressed, and the $4$e peak becomes the dominant feature. The plots in blue and orange are line cuts in the power map taken at $\Phi_{\rm ext} = 0.22~\Phi_0$ and $\Phi_{\rm ext} = \Phi_0/2$ respectively, as indicated by the arrows. In c), we bias the SQUID at $\Phi_{\rm ext} = \Phi_0/2$, and fix $ V_{\rm G2} = -0.875 $~V. We measure the SQUID differential resistance as a function of current bias and $ V_{\text{G1}} $. Moving from left to right, we go from $ I_{\text{c2}} >  I_{\text{c1}}$ to $I_{\text{c1}} >  I_{\text{c2}}$, crossing a balanced configuration. d) Same as in b) but for the gate and flux configuration as in c). For specific values of $ V_{\text{G1}} $, we see a clear increase in the visibility of the $4$e peak. The plots in blue and orange are line cuts in the power map taken at $ V_{\text{G1}} =-0.89$~V and $ V_{\text{G1}} =-0.8$~V respectively. e) Same as in a), but for $ V_{\rm G1} = -0.9 $~V and $ V_{\rm G2} = -1 $~V. In this unbalanced configuration, there is diode effect. f) Same as in b) but for the gate configuration as in e). Here, throughout the flux bias range, the $ 2 $e peak remains the dominant feature.}
	\label{fig:Fig3}
\end{figure*}
In the following, we characterize the dc and ac Josephson effect of the individual JJs. To this end, we measure the gate dependence of the critical current and the radiation spectrum of each junction, while the neighbouring one is fully depleted.
Figure~\ref{fig:Fig2}a) and Fig.~\ref{fig:Fig2}b) show the differential resistance of JJ$ _1 $ and JJ$ _2 $ as a function of current bias $I$ for different gate-voltages. We identify the critical current $I_{\rm c}$ as the boundary between the superconducting regime (dark blue) and the ohmic regime (turquoise). At negative gate voltages ($ V_{\rm G1}\leq -0.9$\,V and $ V_{\rm G2}\leq -1.5$\,V ) $ I_{\rm c} $ is negligibly small, but it can be gradually increased with increasing $ V_{\rm Gi}$. $I_{\rm c,max} $ saturates to $I_{\rm c1,max}=\SI{1.1}{\micro\ampere}$ for JJ$ _1 $ and $I_{\rm c2,max}=\SI{0.8}{\micro\ampere}$  for JJ$ _2 $ at around $ V_{\rm Gi}=0.5$\,V. The slight differences in the gate dependence of the two junctions is attributed to a different junction width and gate geometry. To estimate the $I_{\rm c}R_{\rm n} $ product of the junctions, we measure the resistance at voltage bias larger than twice the superconducting gap of the leads as obtained from multiple Andreev reflection measurements conducted on a different chip of the same wafer. Subtracting the shunt resistor, we obtain a normal state resistance of the junction  $R_{\rm n}\sim\SI{90}{\ohm} $, corresponding to a $I_{\rm c}R_{\rm n}\sim\SI{90}{\micro\volt}$. We also note that potential errors in estimating $R_{\rm n}$ might have led to an underestimation of the $I_{\rm c}R_{\rm n}$ product. Nonetheless, the significantly large $I_{\rm c}R_{\rm n}$ product indicates a high-quality Josephson junction with a uniform current distribution.

In Fig.~\ref{fig:Fig2}c) and Fig.~\ref{fig:Fig2}d), the $IV$-curves at $ V_{\rm G1}=-0.75$~V and $ V_{\rm G2}=-0.7$~V, respectively are obtained by integrating the measured $dV/dI$ curves along the white dashed lines in Fig.~\ref{fig:Fig2}a) and Fig.~\ref{fig:Fig2}b). Both junctions show an ohmic behaviour down to $2~\mu$V, which allows stable voltage biasing in the microwave regime of the Josephson emission. 

According to the ac Josephson effect, the phase difference of a voltage biased Jospehson junction will evolve linearly in time following
\begin{equation}
\varphi(t)~=~\frac{2\pi}{\Phi_0}Vt,
\label{eqn:phase_evolution}
\end{equation}
with $ V $ being the voltage drop across the junction. Consequently, an applied dc voltage causes a oscillating supercurrent at the Josephson frequency $f_{\rm J}=2\text{e}V/h$. This transforms into the emission of microwave photons at $ f_{\rm J} $. If higher harmonics are present, photon emission at higher frequencies $f_{\mathrm{J},m}=m\times2\text{e}V/h$ also occurs~\cite{Basset2019,haller2023ac}. In Fig.~\ref{fig:Fig2}e) we show the expected peak evolution in the emission spectrum of voltage biased JJ as a function of detection frequency $f_{\rm det}$ and $V$. For every voltage bias position, peaks emerge in the emission spectrum, if the detection frequency matches an integer multiple of the Josephson frequency $f_{\rm det}=f_{\mathrm{J},m}$. These peaks induce a fan-like pattern, capturing the linear relation between voltage and the emission frequency with slope $h/(m2\text{e})$. Emission lines evolving as $hf_{\rm det}/(m2\text{e})$ correspond to the coherent transport of $ m $ Cooper-pairs across the junction (red, orange, and pink dashed lines for $ m=$~1, 2, and 3). In addition to the fan-like pattern,  replicas of the Josephson emission lines can appear at a constant frequency offset on the right and on the left of the predicted peak position due to photon-assisted emission through environmental modes~\cite{Haller_thesis}. A photon from a spurious environmental mode can be upconverted to a detector photon by taking up the energy $2\text{e}V$ provided by the inelastic tunnelling of a Cooper-pair (right shift in frequency). The energy balance in this case reads $hf_{\rm det}=hf_{\rm env}+2\text{e}V$, where $f_{\rm env}$ corresponds to the resonant frequency of an environmental cavity. Such resonance can be caused for example by a standing wave pattern along the microwave lines.  The complementary process is also possible, meaning that a photon coming from a Cooper-pair tunnelling can be downconverted to a detector photon by giving up the energy $ hf_{\rm env}$ to the environment (left shift in frequency). The energy balance in this case reads  $hf_{\rm det}= 2\text{e}V - hf_{\rm env}$.

In Fig.~\ref{fig:Fig2}f) and Fig.~\ref{fig:Fig2}g) we plot the normalized radiation power $P _{\rm det,norm} $ as a function of $f_{\rm det}$ and $V_{\rm int}$ for JJ$ _1 $ and JJ$ _2 $ respectively. The power is normalized at each detection frequency to compensate for the frequency-dependent background. A pronounced emission peak at frequency $f_{J,1}$ (red dashed line) corresponding to the 2e single Cooper-pair transport is measured over the entire frequency range from $ 5 $~GHz to $ 8 $~GHz. The signal due to the 4e double Cooper-pair transport at frequency $f_{J,2}$ (orange dashed line) is weaker but becomes clearly visible in the emission spectrum around $7$~GHz (orange arrow). Emission peaks at frequencies corresponding to higher harmonics, $ m>2 $, are below our detection limit. In addition to the fan-like pattern, there is a strong replica of the fundamental Josephson emission appearing at a constant frequency offset ($f_{\rm env} \sim 1.95$~GHz) on the right of the predicted peak position. Its contribution diminishes for $f_{\rm det}>6$~GHz. Changes in power spectrum as a function of detection frequency arise from a frequency-dependent probability of photon emission due to inelastic Cooper pair tunnelling. This emission probability depends on the impedance of the environment surrounding the Josephson junction~\cite{Jebari2018}, which in turns has a complex behaviour as a function of frequency caused, for example, by standing wave patterns in the rf lines due to spurious impedance mismatch conditions. By setting $f_{\rm det}=7.1$~GHz, we can disregard the contribution of this environmental mode in the following investigation.
%-----------------------------------------------------------------------------------------------------------
\subsection{Ac and dc Josephson effect from a SQUID}
\label{subsec:Radiation from a SQUID}
\vspace{-0.3cm}
%-------------------------------------------------------------------------------------------------------------------
Next, we exploit the interference between the two junctions when both carry a finite supercurrent in order to realize an effective Josephson element with negligible first harmonic component. We require two conditions: (i) the flux is to be set to $\Phi_{\rm ext}=\Phi_0 / 2$, and (ii), the JJs are gate-tuned into balance, such that $c_{\rm 1,\rm JJ_1}=c_{1,\rm JJ_2}$. The key challenge in the experiment is the balancing of the junctions. As a solution, we adopt an approach proposed in~\cite{Souto2022} that is based on the observation that $ I_{\rm c}$ for the forward and reverse current-bias directions, $I_{c,+}$ and $I_{c,-}$, is mismatched, unless both junctions are balanced and $\Phi_{\rm ext}=n\Phi_0 / 2$ with n being an integer. To balance the SQUID, we look at regions in gate voltage without diode effect, meaning $I_{c,+}$ and $I_{c,-}$ are equal (symmetric junctions).

In Fig.~\ref{fig:Fig3} we measure the SQUID in three different configurations. Firstly, we fix the gate voltages such that the junctions are symmetric and sweep $ \Phi_{\rm ext} $. Secondly, we fix $V_{\text{G2}}$ and sweep $V_{\text{G1}}$ at $\Phi_{\rm ext}=\Phi_0 / 2$. Finally, we fix the gate voltages and sweep $ \Phi_{\rm ext} $ in the case of asymmetric junctions.

In Fig.~\ref{fig:Fig3}a) we plot the SQUID differential resistance $dV/dI$ as a function of current bias $I$ and $ \Phi_{\rm ext} $ in a gate configuration where $I_{\rm c1} \approx I_{\rm c2}$. No diode effect is observed over the entire flux bias range. Differences between the gate values at which symmetry is achieved and those expected from Fig.~\ref{fig:Fig2}a) and Fig.~\ref{fig:Fig2}b) are caused by the fact that the critical current of each junction depends on weather the junction is measured individually or embedded in SQUID~\cite{haxell2023}. Simultaneously, we measure the SQUID ac emission at fixed detection frequency $f_{\rm det}=7.1$~GHz. Figure~\ref{fig:Fig3}b) shows the normalized
radiation power $P _{\rm det,norm} $ as a function of $ \Phi_{\rm ext} $ and integrated voltage drop over the SQUID $ V_{\rm int}$. Because the signal peaks at $V_{m}=hf_{\rm det}/(m2\text{e})$, we scale the voltage axis by $hf_{\rm det}/2\text{e}$. The emission pattern changes in a striking manner around  $\Phi_{\rm ext} = \Phi_0/2$. The fundamental Josephson signal at a scaled $V_{\rm int}=1$, corresponding to the $2$e supercurrent, vanishes almost completely, while a sharp bright peak at a scaled $V_{\rm int}=1/2$ appears, that corresponds to the radiation signal coming from the simultaneous inelastic transport of pairs of Cooper-pairs. An additional horizontal line is visible in the map due to the spurious environmental mode, as addressed before. On the right panels, we plot cuts along $V_{\rm int}$ at $\Phi_{\rm ext} = 0.22~\Phi_0$ (blue) and $\Phi_{\rm ext} = \Phi_0/2$ (orange). The radiation power is here presented in a linear scale. At $\Phi_{\rm ext} = \Phi_0/2$, the 4e peak emerges as the dominant feature, yet its amplitude is approximately $ \sim 25 $ times smaller compared to the amplitude of the 2e peak measured at $\Phi_{\rm ext} = 0.22~\Phi_0$. This is expected, since the amplitude of the power emission peak is proportional to the square of $ I_{\rm c} $, which at $\Phi_{\rm ext} = \Phi_0/2$ is only determined by the second harmonic of the CPR, and is $ \sim 5 $ times smaller than $ I_{\rm c} $ at $\Phi_{\rm ext} = 0.22~\Phi_0$. A detailed analysis of the ratio between the 4e and 2e peaks can be found in the Supplementary Discussion 1.
%~\hyperref[sec:Evaluation of Fourier Components dependence on External Flux]{1}

We investigate the dependence of the emission spectrum as a function of $V_{\text{G1}}$, when the magnetic flux is set to $\Phi_{\rm ext}=\Phi_{0}/2$ and $V_{\text{G2}}= -0.875$~V, shown in Fig.~\ref{fig:Fig3}c) and d). Away from the balanced configuration, the more distinct peak in the emission spectrum is the one corresponding to the $2$e transport. However, once we approach the balanced situation at $V_{\text{G1}}\sim -0.8$~V the signal at $V_{1}=hf_{\text{det}}/2\text{e}$ is suppressed, and instead, the dominant peak in the emission spectrum becomes the one at $V_{2}=hf_{\text{det}}/4\text{e}$.
%In Fig.~\ref{fig:Fig3}e) we notice a finite diode effect, meaning that the flux was not perfectly centred around $\Phi_{0}/2$. This is also observed in the radiation map in Fig.~\ref{fig:Fig3}f) as a non perfect cancelling of the 2e peak. However, the 4e peak becomes the dominant feature for several values of $V_{\text{G1}}$.

Lastly, in Fig.~\ref{fig:Fig3}e) we plot the $dV/dI$ of the SQUID as a function of $I$ and $ \Phi_{\rm ext} $ in a gate configuration where $I_{\rm c1} \not= I_{\rm c2}$. Apart from $\Phi_{\rm ext} = \Phi_0/2$, there is a clearly visible diode effect. Figure~\ref{fig:Fig3}f) shows $P _{\rm det,norm} $ as a function of $\Phi_{\rm ext}$ in the same gate configuration. The 2e emission peak remains the dominant feature throughout the whole flux bias range. Its amplitude decreases asymmetrically on the left- and right-hand side of $\Phi_{\rm ext} = \Phi_0/2$, following the asymmetry of the SQUID critical current. Even though the junctions are not balanced, one can still see that the emission signal slightly increases at voltages $V_2 = hf_{\rm det}/4\text{e}$, in the vicinity of $\Phi_{\rm ext} = \Phi_0/2$.

These findings show that a continuous transition between a $2$e and a $4$e supercurrent can be achieved by tuning both gate voltages and the magnetic flux. Importantly, the 4e supercurrent dominates over a finite window in parameter space and is not limited to exactly matching boundary conditions.

%-------------------------------------------------------------------------------------------------------------------
\subsection{Shapiro steps}
\label{subsec:Shapiro_steps}
\vspace{-0.3cm}
%-------------------------------------------------------------------------------------------------------------------

\begin{figure}[!b]
	\centering
	\includegraphics[width=\columnwidth]{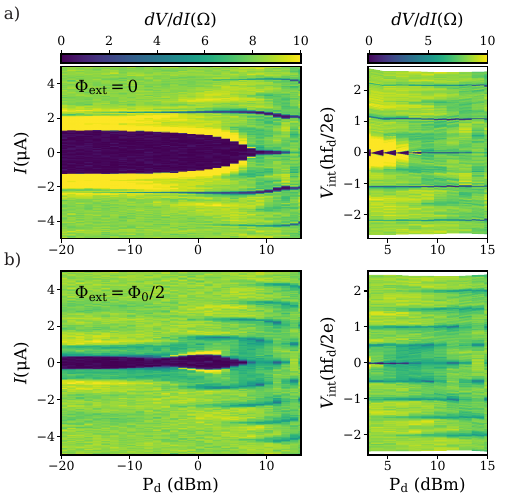}
	\caption{\textbf{Shapiro steps measurements at zero and half flux quantum for symmetric junctions.} a) On the left, differential resistance $ dV/dI $ as a function of drive power $ P_{\text{d}} $ and bias current $ I $ at constant drive frequency $ f_{\text{d}} = 7.5$~GHz and zero external flux $ \Phi_{\rm ext} =0$ for $ V_{\text{G1}}=-0.73$~V and $ V_{\text{G2}}=-0.5$~V. The drops in differential resistance correspond to the emergence of Shapiro steps. On the right, differential resistance as a function of $ P_{\text{d}} $ plotted vs normalized voltage drop $V_{\text{int}}$ over the SQUID. At zero flux, mostly integer Shapiro steps are visible. b) Same as in a), but at $ \Phi_{\rm ext} =\Phi_{\text{0}}/2$. The destructive interference of the first harmonics produces a CPR with double the periodicity of the individual junctions, inducing the emergence of half integer Shapiro steps.}
	\label{fig:Fig4v2}
\end{figure}
So far, we have used the Josephson radiation measurements to identify the emergence of a $4$e supercurrent in the SQUID. In the last part of this work, we discuss Shapiro step measurements that complement the radiation experiment. When a microwave drive tone is sent to a JJ, distinct voltage plateaus in the $V(I)$ characteristic appear, known as Shapiro steps\cite{SHAPIRO1964,Raes2020,Ueda2020,Iorio2023,valentini2023radio}. For a conventional $ \sin(\varphi) $ junction, each plateau corresponds to a Cooper-pair absorbing $ n $ photons with frequency $f_{\rm d}$ to overcome the Shapiro step voltage $V_n$, and the energy relation reads $2\text{e}V_n = nhf_{\rm d}$. The presence of higher harmonics in the CPR of the junction changes the energy relation to $2meV_n = nhf_{\rm d}$, corresponding to $ m $ Copper-pairs absorbing $ n $ photons to overcome the voltage step.

We apply a microwave tone of fixed frequency $f_{\textrm{d}} = 7.5$~GHz to the SQUID with different output power $P_{\rm d}$ values. The signal is applied to the microwave input line, connecting the device to the amplification chain through a directional coupler (see Supplementary Note 1). In Fig.~\ref{fig:Fig4v2}(a), we plot the SQUID differential resistance $dV/dI$ at $\Phi_{\rm ext}=0$ as a function of current bias $I$ and $P_{\rm d}$ in a symmetric gate configuration. In the left panel we plot $dV/dI$ versus $I$, and on the right we plot the data as a function of the integrated voltage $V_{\rm int}$ scaled by $hf_{\rm d}/2\text{e}$. Shapiro-steps occur at integer values of the scaled voltage as dips in differential resistance. 

The data in Fig.~\ref{fig:Fig4v2}b) is measured for the same gate values as in Fig.~\ref{fig:Fig4v2}a), but at $\Phi_{\rm ext}=\Phi_{0}/2$. In this configuration, the SQUID resembles an effective $ \sin(2\varphi) $ junction because the 2e supercurrent is suppressed. The energy relation for the appearance of Shapiro steps is given in this case by $4\text{e}V_n = nhf_{\rm d}$, resulting in a doubling of the number of observed steps. In line with the theoretical expectations~\cite{Souto2022}, both integer and half-integer Shapiro steps are equally visible in the data. Differences between this measurement and Shapiro steps measurements performed on Josephson junction with a high quality factor~\cite{LeCalvez2019} are attributed to the $ 10~\si{\ohm} $-shunt in our device.

%-------------------------------------------------------------------------------------------------------------------
\section{Conclusion}
\label{sec:Conclusion}
\vspace{-0.3cm}
%-------------------------------------------------------------------------------------------------------------------
We have demonstrated the realization of an effective $ \sin(2\varphi) $ Josephson junction using a dc SQUID consisting of two planar Josephson junctions formed in a proximitized InAs two-dimensional electron gas. We probe the emergence of a dominant second harmonic in the CPR of the SQUID by measuring the ac Josephson effect as a function of gate voltages and magnetic flux. Photon emission at the fundamental Josephson frequency is suppressed when the SQUID is in a symmetric configuration and biased at half flux and instead, photons are only emitted at $ f_{\rm J,2} $. We provide evidence on how to continuously tune from the $2\text{e}$ to the $4\text{e}$ supercurrent regime by adjusting the junction gate voltages and the external magnetic flux. The results are further substantiated through complementary Shapiro step measurements in a symmetric SQUID configuration at half flux, revealing additional half-integer steps with same visibility as the integer steps.

Our results indicate, that a robust $ \sin(2\varphi) $ JJ can be engineered and could used to realize parity-protected qubits with this material system. Such parity-protected qubit provides an alternative route to the protection of quantum information in superconducting devices and may complement alternative approaches based on fluxonium qubits \cite{manucharyan2009fluxonium,nguyen2019high,hazard2019nanowire,somoroff2021millisecond} and qubits based on topological wavefunctions \cite{hoffman2016universal,ranvcic2019entangling,schrade2018majorana,schrade2022quantum,landau2016towards,plugge2017majorana,karzig2017scalable}.
Looking ahead, the 2D platform would make it easier to further protect the qubit from noise and offsets by concatenating several SQUIDs in parallel~\cite{Schrade2022}.

\section{Methods}
\label{sec:Methods}
\vspace{-0.3cm}
%-------------------------------------------------------------------------------------------------------------------
The proximitized InAs two-dimensional electron gas used in this project is grown starting from a semi-insulating InP (100) substrate. A $ \SI{1}{\micro\meter} $ thick  In$ _x $Al$ _{1-x} $As buffer layer is used to match the lattice constant of InP to the one of InAs. The quantum well consists of a $ 7 $~nm InAs layer sandwiched between a $ 10 $~nm (top barrier) and a $ 4 $~nm (bottom barrier) In$ _{0.75} $Ga$ _{0.25} $As layer. The $ 10 $~nm Al layer is epitaxially grown on top of a capping GaAs thin film without breaking the vacuum, ensuring a pristine interface between the semiconductor and the superconductor. Here we show results obtained from a wafer stack with mobility $ \mu $ = 11,000 cm$ ^2 $V$ ^{-1} $s$ ^{-1} $ at electron densities of 2.0 $ \times $ 10$ ^{12} $cm$ ^{-2} $, measured on a different chip coming from the same wafer.

The device is fabricated using standard electron beam lithography techniques. The SQUID is electrically isolated by etching the Al layer and $ 300 $~nm of buffer around it. First, the Al film is removed with Al etchant Transene D, followed by a deep III–V chemical wet etch H$ _2 $O:C$ _6 $H$ _8 $O$ _7 $:H$ _3 $PO$ _4 $:H$ _2 $O$ _2 $ (220:55:3:3). Next, JJs are formed by selectively removing the Al over 150nm-long stripes on each branch of the loop. A 15 nm-thick layer of insulating HfO$ _2 $ is grown by atomic layer deposition at a temperature of 90 °C over the entire sample. The set of gates are realized in two steps. A thin Ti/Au (5/25 nm) layer is evaporated on top of the mesa to define the gate geometry, and then leads and bonding pads are defined by evaporating a Ti/Au (5/85 nm) layer at and angle of $ \pm $17° to overcome the mesa step. More information about the full wafer stack and the fabrication procedure can be found in Ref.~\cite{Shabani2014,Kjaergaard2016,Shabani2016,Lee2019}.

\section{Data Availability}
\label{sec:Data Availability}
\vspace{-0.3cm}
%-------------------------------------------------------------------------------------------------------------------
All data in this publication is available in numerical form at: \url{https://doi.org/10.5281/zenodo.7969736}.

\section{Authors Contributions}
\label{sec:Authors Contributions}
\vspace{-0.3cm}
%-------------------------------------------------------------------------------------------------------------------
C.C. fabricated the device and performed the measurements with the help of R.H.. C.C. and R.H. analysed the data with inputs from A.C.C.D. and C.S.. C.C. wrote the manuscript with inputs from all authors. C.S. initiated the project. T.L. and M.M. provided the InAs material.

\section{Competing Interests}
\label{sec:Competing Interests}
\vspace{-0.3cm}
%-------------------------------------------------------------------------------------------------------------------
The authors declare no competing interests.

\begin{acknowledgments}
	\vspace{-0.3cm}
	We thank  C. M. Marcus for his support in initiating this work and collaboration. We thank Libin Wang for his help in the development of the fabrication procedure. We thank Joost Ridderbos and Gerg\H{o} F\"{u}l\"{o}p for their help with the setup and the understanding of the measurements in the early stage of the experiment. We thank Martin Endres for fruitful discussions. This research was supported by the Swiss Nanoscience Institute (SNI), the Swiss National Science Foundation through grants No 172638 and 192027, and the QuantEra project SuperTop. We further acknowledge funding from the European Union’s Horizon 2020 research and innovation programme, specifically a) from the European Research Council (ERC) grant agreement No 787414, ERC-Adv TopSupra, b) grant agreement No 828948, FET-open project AndQC, and c) grant agreement 847471, project COFUND-QUSTEC. Constantin Schrade acknowledges support from the Microsoft Corporation.
\end{acknowledgments}

\newpage
%%%%%%%%%%%%%%%%%%%%%%%%%%%%%%%%%%%%%%%%%%%%%%%%%%%%%%%%%%%%%%%%%%%%
% Include bibtex file here with references
%%%%%%%%%%%%%%%%%%%%%%%%%%%%%%%%%%%%%%%%%%%%%%%%%%%%%%%%%%%%%%%%%%%%
\renewcommand{\bibsection}{\section*{References}} %

%apsrev4-2.bst 2019-01-14 (MD) hand-edited version of apsrev4-1.bst
%Control: key (0)
%Control: author (8) initials jnrlst
%Control: editor formatted (1) identically to author
%Control: production of article title (0) allowed
%Control: page (0) single
%Control: year (1) truncated
%Control: production of eprint (0) enabled
%

%\bibliography{bib}
\clearpage
\widetext

\setcounter{equation}{0}
\setcounter{figure}{0}
\setcounter{table}{0}
\setcounter{page}{1}
\setcounter{section}{0}

\renewcommand{\thefigure}{S\arabic{figure}}
\renewcommand{\theequation}{S\arabic{equation}}
\renewcommand{\thesection}{S\Roman{section}}
\renewcommand{\bibnumfmt}[1]{[S#1]}
\renewcommand{\citenumfont}[1]{S#1}

\textbf{\centering\large Supplementary Material:\\}
\textbf{\centering\large Charge-4e supercurrent in an InAs-Al superconductor-semiconductor heterostructure\\}
\vspace{1em}
\section{Supplementary Note 1: Extended Setup Description}
\label{sec:Extended Setup Description}
\vspace{-0.3cm}

\begin{figure}[htb]
	\centering
	\includegraphics[width=\columnwidth]{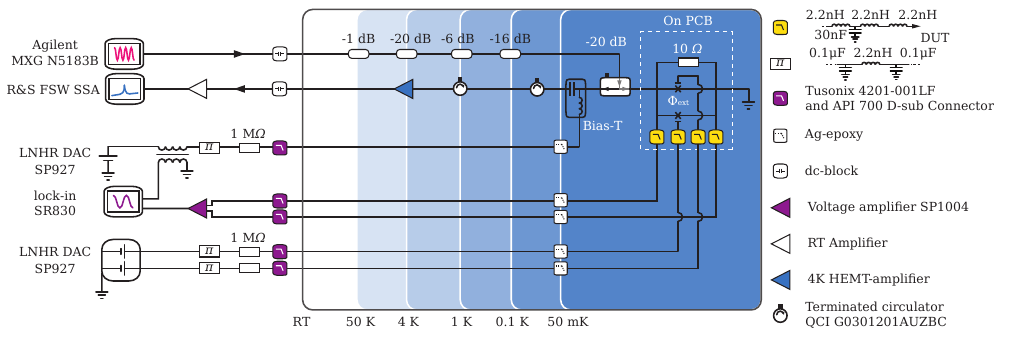}
	\caption{Detailed description of the measurement setup.}
	\label{fig:s1}
\end{figure}
The chip is mounted on a double-sided Ni/Au plated Roger\textsuperscript{\textcopyright} 4350 PCB screwed onto a copper plate. A $10\Omega$ metal film resistor is soldered on the back side of the PCB in-between the ground plane and the central conductor of the SMP connector. The device is glued onto the central copper area such that the microwave bond terminals are as close as possible to the PCB counterpart. The PCB central conductor is bonded to one side of the SQUID, whereas the other side is bonded to the PCB ground resulting in a resistively shunted junction configuration. Both wire bonds are made as short as possible. Additional bond wires connect to the differential voltage measure and gate voltage supplies.

The dc supply lines are filtered at room temperature, on the PCB, and are thermalized to the mixing chamber plate via silver-epoxy filters that provide a cut-off of $ \approx 6$~MHz. A current bias is generated by a $1$~M$\Omega$ resistor in series with a voltage source. The current couples via a bias-tee to the microwave line that connects through the device to ground. The voltage drop across the junction is measured differentially with a voltage amplifier and lock-in techniques.

A constant voltage drop across a Josephson junction leads to an oscillating current. The amplification chain collects this radiation signal and feeds it to a R\&S FSW SSA spectrum analyser. The ac signal is coupled via the bias-tee to a cryogenic HEMT amplifier (nominally $+40$~dB gain) located on the 4K stage that is isolated from the device with two terminated circulators. The signal is further amplified with a room temperature amplifier (nominally $+40$~dB gain). The following measurement parameters are set at the spectrum analyzer to sense the amplified Josephson emission: detection bandwidth $20$~MHz, span $22$~MHz, resolution bandwidth $20$~MHz, video bandwidth $100$~Hz and 1001 points resulting in a sweep time of $1$~s. In addition to the sensing line, a drive line connects to the device via a directional coupler. A Agilent MXG N5183B signal generator is used to send an microwave tone to the SQUID tp o perform Shapiro step measurements. The detection bandwidth is limited by the directional coupler to $2.5 - 8.5$~GHz.

The external magnetic field flux $ \Phi_{\rm ext} $ is applied with a 3-axis vector magnet sourced with a Keithley $ 2400 $.

Measurements presented in the following are performed in a dilution refrigerator with a base temperature of $\sim 50\,{\textrm{mK}}$.
\newpage
\section{Supplementary Discussion 1: Peaks amplitude dependence on External Flux}
\label{sec:Evaluation of Fourier Components dependence on External Flux}
\vspace{-0.3cm}

\begin{figure}[htb]
	\centering
	\includegraphics[width=\columnwidth]{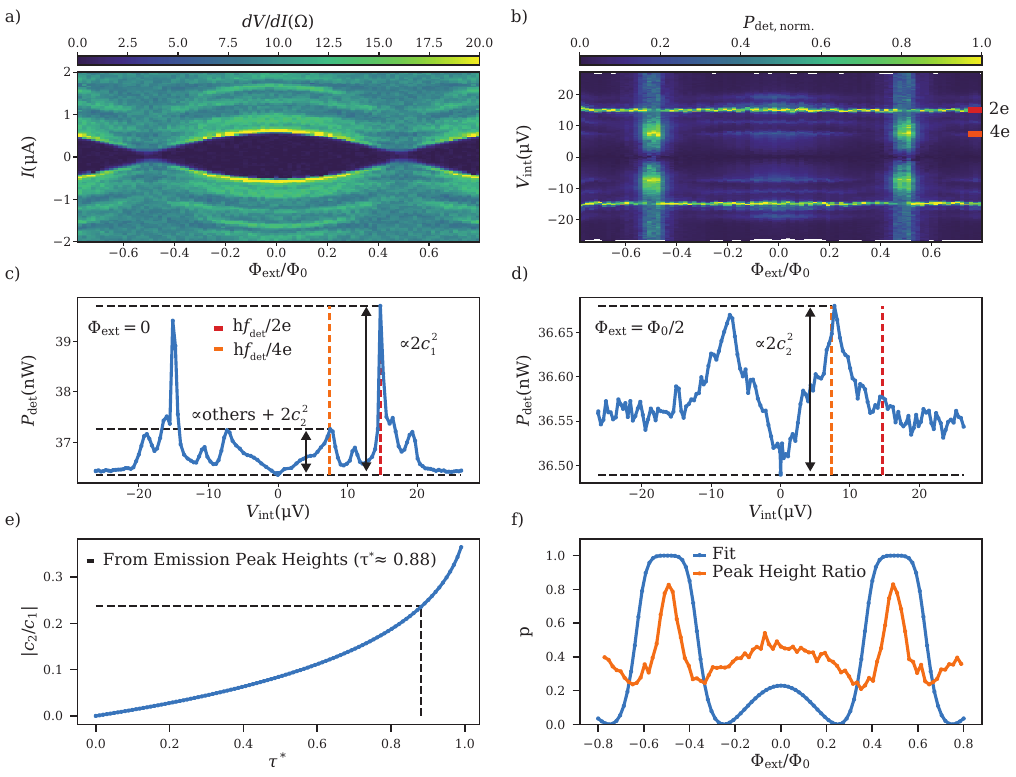}
	\caption{a) Differential resistance $ dV/dI $ of the SQUID as a function of external flux $ \Phi_{\mathrm{ext}} $ and current bias $ I $ for symmetric junctions. Here, $ V_{\rm G1} = -0.865 $~V and $ V_{\rm G2} = -0.845 $~V. b) Normalized radiation power $P _{\text{det,norm.}}$ at $ f_{\text{det}} = 7.1~$GHz as a function of external flux $\Phi_{\mathrm{ext}}$ and normalized voltage drop over the SQUID. The map is measured as the same time as in a). We show the periodic suppression of the 2e peak and the emergence of a dominant 4e peak. c) Cut in the emission spectrum in b) at zero flux. The horizontal black dashed lines highlight, from bottom to top, the signal baseline, the 4e and the 2e peak height. d) Cut in the emission spectrum in b) at half flux. The horizontal black dashed lines highlight, the signal baseline and the 4e peak height. A small 2e emission signal is visible at $V_{\rm int} = -hf_{\rm det}/2\text{e}$. e) Absolute value of the ratio between the second and first Fourier coefficients $ \left|\mathit{c}_{\mathrm{2}}/\mathit{c}_{\mathrm{1}}\right| $ as a function of the junction transparency $ \tau $. The dashed line corresponds to the square root of the ratio between the 4e and the 2e emission peak amplitude at half and zero flux respectively. By comparing this with the $ \left|\mathit{c}_{\mathrm{2}}/\mathit{c}_{\mathrm{1}}\right| $ ratio calculated from the fit, we extract a junction transparency of $ 0.88 $. f) The ratio $ p $ (see main text) as a function of $\Phi_{\mathrm{ext}}$ extracted from b). We compare this to the evolution of the ratio $ \left|\mathit{c}_2\right|/\sqrt{\mathit{c}_1^2+\mathit{c}_2^2} $ as a function of $\Phi_{\mathrm{ext}}$ extracted from the Fourier fit of the CPR of a SQUID formed by two identical single channels junctions with transparency 0.88.}
	\label{fig:s3}
\end{figure}

In this sections, we attempt to relate the ratio between the 4e and 2e emission peaks amplitude to the harmonics content of a junction with effective transparency $ \tau^* $. Considering a Josepshon junction as an ac current source in parallel with a resistor $ R_{\rm s} $\cite{Haller_thesis}, the power dissipated in the circuit is given by:
\begin{equation}
P = \frac{R_{\rm s}I_{\rm c}^2}{2} = \frac{R_{\rm s}}{2}\sum_{m=1}^{\infty}c_m^2(\tau^*),
\label{eq:dissipated_power}
\end{equation}
where $ I_{\rm c} $ is the critical current of the junction and $ c_m^2(\tau^*) $ is the contribution to the dissipated power coming from the $ m $-th harmonic.
%Since part of this power will be dissipated across the $ 50~\Omega $-termination of the spectrum analyser, Fourier coefficients of the junction CPR are proportional to the square root of the amplitudes of the emission peaks in the power spectrum.
For each detection frequency, we define the amplitude of the $ m $-th peak $ P_{\mathrm{det,}m} $ as the difference between the peak height and the minimum of the detected power. The ratio of the Fourier coefficients of the CPR of the junction is directly proportional to the square root of the amplitudes of the emission peaks:
\begin{equation}
\left|\frac{\mathit{c}_{\mathrm{2}}}{\mathit{c}_{\mathrm{1}}}\right| \propto \sqrt{\frac{P_{\mathrm{det,}2}}{P_{\mathrm{det,}1}}}.
\label{eq:power_vs_CPR}
\end{equation}
In practice, additional processes increase the amplitude of higher harmonics peaks compared to the one expected from the harmonic content of the CPR~\cite{gross2016applied}. At current bias values $ I $ larger but close $ I_{\rm c} $, part of the current has to flow as normal current through the shunt resistor and displacement current through the parallel capacitance. The interplay between super-, normal and displacement current results in a time varying voltage over the junction that causes emission at multiples of the fundamental Josephson frequency even in the case of a fully sinusoidal CPR~\cite{haller2023ac}. For $ I\gg I_{\rm c} $ instead, most of the current flows as normal current, resulting in a more constant junction voltage. Emission of photons at current bias values such that the ratio $ k = I/I_{\rm c} $ is  $ \gg1 $ is then characterized by a well defined emission frequency.\\

At a detection frequency of $7.1$~GHz, the $4$e emission peak occurs at a voltage bias of $\sim \SI{7.34}{\micro\volt}$. To pinpoint the current value associated with 4e emission, we integrate the measured $dV/dI$ curves across the range of current bias values. Subsequently, we seek the current bias value corresponding to $V_{\mathrm{int}} = \SI{7.34}{\micro\volt}$. Figure\ref{fig:s3}a) shows the differential resistance $dV/dI$ as a function of bias current $I$ and external flux $ \Phi_{\rm ext} $ at $ V_{\rm G1} = -0.865 $~V and $ V_{\rm G2} = -0.845 $~V. In this gate voltage configuration, the two junctions are symmetric. The difference in gate positions at which balance is achieved in this case compared to the one shown in the main text can be attributed to gate hysteresis and gate drift occurring during the time period between measurements. At $ \Phi_{\rm ext} =0$, the critical current of the SQUID  is $ I_{\rm c}\sim 0.7~\mu $A and $ k $ is only $ \sim 1.6$. This results in a time varying voltage over the SQUID, and therefore in a finite contribution to the 4e emission amplitude that adds on top of the one coming from the CPR of the junctions. On the other hand, at $ \Phi_{\rm ext} =\Phi_{0}/2$ the critical current of the SQUID is $ I_{\rm c}\sim 0.12~\mu $A, corresponding to a much higher current bias to critical current ratio for the 4e emission $k \sim 9.5$. The lower critical current value also reduces the contribution from environmental modes, so that at $ \Phi_{\rm ext} =\Phi_{0}/2$, the 4e peak amplitude can be more faithfully related to the harmonic content of the CPR. Since each junction contribute $ c_{\mathrm{i}}^2 $ to the emission signal, the peak amplitudes for a SQUID in the symmetric configuration are proportional to $ 2c_{\mathrm{i}}^2 $.\\

Figure~\ref{fig:s3}b) shows the normalized radiation power  $ P_{\mathrm{det,norm.}} $ measured at the same time as in Fig.\ref{fig:s3}a) at fixed detection frequency $f_{\rm det}=7.1$~GHz as a function of integrated voltage $ V_{\mathrm{int}} $ and $ \Phi_{\rm ext} $. One can observe a periodic suppression of the 2e peak and emergence of the 4e peak.\\

In Fig.\ref{fig:s3}e), we plot the second to first harmonic ratio $ \left|\mathit{c}_{\mathrm{2}}/\mathit{c}_{\mathrm{1}}\right| $ of a non-sinusoidal CPR as a function of the junction transparency $ \tau^* $. The Fourier coefficient are extracted for each  $ \tau^* $ by fitting the CPR with a Fourier series with 10 harmonics. The $\left|\mathit{c}_{\mathrm{2}}/\mathit{c}_{\mathrm{1}}\right|$ ratio calculated from Eq.\eqref{eq:power_vs_CPR} using as $ P_{\mathrm{det,}1} $ the 2e peak emission amplitude at $ \Phi_{\rm ext} =0$, and as $ P_{\mathrm{det,}2} $ the 4e peak emission amplitude at $ \Phi_{\rm ext} =\Phi_{0}/2$, corresponds to an effective transparency of $ \tau^*\sim 0.88 $. This is in good agreement with the effective transparency extracted for the same device using an asymmetric SQUID configuration\cite{ciaccia2023gate}.\\

Finally, Fig.\ref{fig:s3}f) shows the normalized ratio $p = \sqrt{P_{\mathrm{det,}2}}/\sqrt{P_{\mathrm{det,}1}^2+P_{\mathrm{det,}2}^2} $ as a function of $\Phi_{\mathrm{ext}}$ as extracted from Fig.\ref{fig:s3}b). In the same figure we plot the evolution of the ratio $ \left|\mathit{c}_2\right|/\sqrt{\mathit{c}_1^2+\mathit{c}_2^2} $  with $\Phi_{\mathrm{ext}}$ as extracted from a fit of Fourier series to the CPR of a symmetric SQUID with effective transparency $ \tau^* = 0.88 $. The ratio extracted from amplitudes of the peaks is vertically shifted relative to the fit. We attribute this shift to additional contributions to the 4e peak originating from a time-varying voltage across the SQUID and environmental processes.

\newpage
\section{Supplementary Discussion 2: Evolution of the 4e Peak amplitude with Gate voltages Configuration}
\label{sec:Evaluation of Fourier Components dependence on the Gate voltages Configuration}
\vspace{-0.3cm}
\begin{figure}[htb]
	\centering
	\includegraphics[width=\columnwidth]{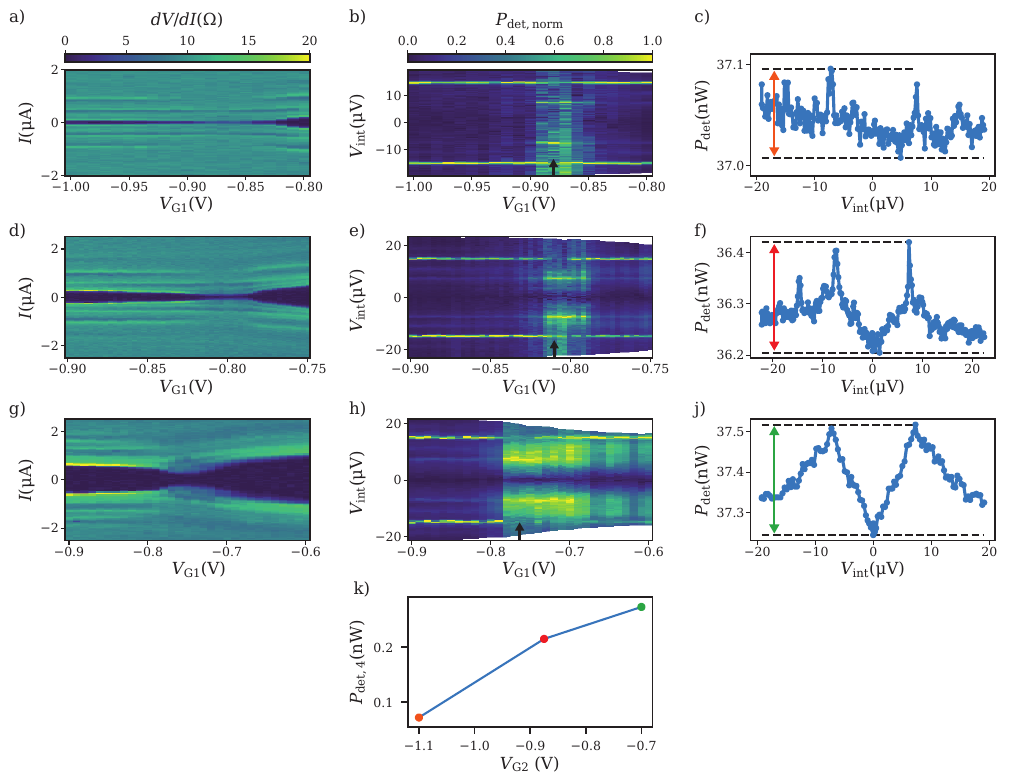}
	\caption{a) Differential resistance $ dV/dI $ of the SQUID as a function of gate voltage $ V_{\rm G1} $ and current bias $I$. The external flux is kept at half flux quantum $ \Phi_0/2 $ and $ V_{\rm G2} = \SI{-1.1}{\volt}$. b) Normalized radiation power $P _{\text{det,norm.}}$ at $ f_{\text{det}} = 7.1~$GHz plotted as a function of $ V_{\rm G1} $ and voltage drop $V_{\rm int}$ over the SQUID. c) Radiation power in linear scale as a function of $V_{\rm int}$ shown for $ V_{\rm G1}= \SI{-0.88}{\volt}$, where the SQUID is symmetric as indicated by the black arrows in b), e) and h). d) and g) Same as in a) but for $ V_{\rm G2} = \SI{-0.875}{\volt}$ and $\SI{-0.7}{\volt}$ respectively. e) and h) Same as in b) but for $ V_{\rm G2} = \SI{-0.875}{\volt}$ and $\SI{-0.7}{\volt}$ respectively. f) and j) Same as in c) but for $ V_{\rm G2} = \SI{-0.875}{\volt}$ and $\SI{-0.7}{\volt}$ respectively.k) Amplitude of the 4e emission peak in the symmetric SQUID configuration as a function of $ V_{\rm G2}$.}
	\label{fig:s4}
\end{figure}

In the following, we fix the external flux at half flux quantum $ \Phi_0/2 $ and we study the evolution of the 4e peak emission amplitude in different gate voltage configurations.\\

% We vary $ V_{\rm G2}$ from  $\SI{-1.1}{\volt}$ to  $\SI{-0.7}{\volt}$, and sweep $ V_{\rm G1}$ to balance the SQUID. By increasing $ V_{\rm G2}$, the critical current of the SQUID in the symmetric configuration also increases, together with the 4e peak emission amplitude.\\
Figure~\ref{fig:s4}a), d) and g) show the differential resistance $dV/dI$ of the SQUID as a function of bias current $I$ and gate voltage $ V_{\rm G1}$ at $ \Phi_0/2 $ for different values of $ V_{\rm G2}$. The symmetry point is identified in each plot by the value of $V_{\rm G1}$ at which the critical current of the SQUID reaches its minimum, indicating the cancellation of the contribution from the first harmonics. By increasing $ V_{\rm G2}$, the critical current of the SQUID in the symmetric configuration also increases.\\

At the same time, we measure the radiation signal from the SQUID. In Fig.~\ref{fig:s4}b), e) and h), we plot the normalized radiation power $P_{\text{det,norm}}$ at fixed detection frequency $ f_{\text{det}} = 7.1~$GHz as a function of  $ V_{\rm G1}$ and voltage drop  $V_{\text{int}}$. In Fig.~\ref{fig:s4}c), f) and j), we show cuts in the radiation map in the symmetric configuration as indicated by the black arrows. The 4e peak amplitude increases with the critical current of the SQUID (see Fig.~\ref{fig:s4}k)) and at the same time, the width of the peak becomes larger. The last can be understood by looking at the $ k $ ratio: decreasing $k$, the voltage drop over the SQUID at the current bias value corresponding to the 4e emission will be less well defined, leading to a broader emission peak.\\

The ability to tune the amplitude of the 4e peak over a wide range is a significant advantage of the 2D platform compared to the nanowire platform for the realization of a parity-protected qubit. To achieve a small dephasing rate, the Josephson energy of the second harmonics in the balanced configuration must be much larger than the island charging energy. This is difficult to achieve with a Josephson junction with only a few conduction channels.

%%%%%%%%%%%%%%%%%%%%%%%%%%%%%%%%%%%%%%%%%%%%%%%%%%%%%%%%%%%%%%%%%%%%
% Include bibtex file here with references
%%%%%%%%%%%%%%%%%%%%%%%%%%%%%%%%%%%%%%%%%%%%%%%%%%%%%%%%%%%%%%%%%%%%
\renewcommand{\bibsection}{\section*{Supplementary References}} %

\end{document}